\documentclass[runningheads]{llncs}

\RequirePackage{etex}

\usepackage{graphicx}
\usepackage{amsmath,amssymb,amsthm}
\usepackage{enumitem}
\usepackage{tikz}
\usetikzlibrary{arrows}
\usetikzlibrary{automata,positioning}
\usepackage{makecell}
\usepackage{url}
\PassOptionsToPackage{bookmarks={false}}{hyperref}
\usepackage{hyperref}
\usepackage{pifont}
\usetikzlibrary{matrix,calc,arrows,decorations.markings}
\usepackage[n,advantage,operators,sets,adversary,landau,probability,notions,logic,ff,mm,primitives,events,complexity,asymptotics,keys]{cryptocode}

\theoremstyle{definition}

\begin{document}
\title{DoS Attacks on Blockchain Ecosystem}

\author{Mayank Raikwar\orcidID{0000-0002-5479-5748} \and
Danilo Gligoroski\orcidID{0000-0002-7078-6139}}
%
%
\institute{Norwegian University of Science and Technology (NTNU)
Trondheim, Norway
\email{\{mayank.raikwar,danilog\}@ntnu.no}}
\maketitle             

\begin{abstract}
Denial of Service (DoS) attacks are a growing threat in network services. The frequency and intensity of DoS attacks are rapidly increasing day by day. The immense financial potential of the Cryptocurrency market is a prevalent target of the DoS attack. The DoS attack events are kept on happening in cryptocurrencies and the blockchain ecosystem. To the best of our knowledge, there has not been any study on the DoS attack on the blockchain ecosystem. In this paper, we identify ten entities in the blockchain ecosystem and we scrutinize the DoS attacks on them. We also present the DoS mitigation techniques applicable to the blockchain services. Additionally, we propose a DoS mitigation technique by the use of verifiable delay function (VDF).

\end{abstract}

\section{Introduction}

Blockchain had brought a paradigm shift in digital innovation and the financial world since the advent of Bitcoin~\cite{Nakamoto_bitcoin}. Today, the cryptocurrency market consists of 5424 cryptocurrencies that all together built a financial market worth around \$1.71 trillion (as of 26 May 2021)~\cite{coinmarketcap}. The immense financial potential of the cryptocurrency market has become a growing concern for the targeted attacks. Some of the well-known attacks in current blockchain systems are selfish mining, blockchain forks, 51\% attack, double spending, Sybil attack, and Denial-of-service attacks~\cite{saad2019exploring}. A Denial-of-service (DoS) attack prevents legitimate user requests and depletes the server's resources. Due to the various configurations and decentralized features of blockchain, many of the attacks are preventable. Nevertheless, DoS attacks, especially its distributed variant (DDoS), are still prominent attacks on cryptocurrencies and blockchain-based applications. 

Due to the increasing intensity and frequency of DoS attacks, it is contemplated as one of the biggest and severe threats for the Internet industries. One of the major DoS attacks was mounted on a DNS server in October 2016, which manifested in a cut of access to major websites, including PayPal, Netflix, and Twitter, for several hours~\cite{woolf2016}. The spectrum of DoS attacks can range from DNS services, cloud providers, IoT devices to the cryptocurrency and blockchain market. Nowadays, the cryptocurrency market is a popular target of DoS attacks, with the motivation of ransom, stealing funds, or business competition. In the past, many works\cite{gupta2017taxonomy,Gasti2013,DOULIGERIS2004643} regarding the detection and prevention of DoS attacks have been carried out. Moreover, DoS/DDoS solutions based on blockchain are an emerging area of research. Applying the most recent advances of cryptographic research for the DoS/DDoS\footnote{Throughout the paper, we use DoS to refer to both DoS and DDoS attacks, unless explicitly mentioned.} problem can open new directions and avenues for addressing this ever-present problem.

In the general context of a DoS attack in blockchain, an adversary usually mounts a DoS attack when the cost of mounting an attack is very low. Therefore, various countermeasures, such as increased block size, increased transaction fees, or limiting transaction size have been proposed for mitigating the attacks. However, most of these countermeasures also force legitimate system users to invest their economic or computational power. This behavior shows a dire need to construct new methods for DoS prevention that do not require extra-economic or computational power of blockchain users. In this paper, we study and review DoS attacks on ten different entities in the blockchain ecosystem and possible mitigation techniques. In addition, we propose DoS mitigation by applying the astonishing functionality of verifiable random function (VDF)~\cite{Boneh2018}.


\subsection{Related Work}
Many DoS attacks have been mounted in the blockchain ecosystem and its services in the past few years. Some of these DoS attacks or threats on cryptocurrencies were disclosed a couple of years after they had been discovered. It requires new techniques to detect and counter the attack. Some of those new blockchain-based DoS mitigation techniques are devised from the decentralized nature and the deployed smart contracts of blockchain ~\cite{Rodrigues2017,Singh2020}. Even different machine learning techniques have been proposed to fight the DoS attack in cryptocurrency~\cite{eduardo2021fighting}.

Specifically for the Bitcoin blockchain (as the blockchain of the most popular and valuable cryptocurrency), several DoS attacks have been mounted~\cite{Vasek2014}, which include mining pools, currency exchanges, eWallets, and financial services. Like most high visibility businesses, mining pools and currency exchanges are the primary DoS  targets, which drives them to buy DDoS protection services such as Incapsula, CloudFlare, or Amazon Cloud. A report from September 2020~\cite{Bitcoin-DoS} revealed that the Bitcoin software implementation had a vulnerability for an uncontrolled memory consumption that was repeatedly used as a DoS vulnerability until it was patched in June 2018. This DoS vulnerability existed in many other branched Bitcoin implementations, including Litecoin and Namecoin.

Another major cryptocurrency, Ethereum~\cite{wood2014} has also suffered from DoS attack~\cite{Atzei2017}. In September 2016, a DoS attack against the Ethereum network was begun by exploiting a flaw in its client node. Furthermore, the same week, another DoS attack was mounted on the processing nodes of Ethereum~\cite{wilcke2016}. A recent disclosure on Ethereum shows that a very cheap DoS attack could have brought down the Ethereum main net due to a bug in Geth Ethereum client~\cite{Ethereum-DoS}.

Recent work shows an Incentive-based blockchain denial of service attack (BDoS) ~\cite{Mirkin2020} on Proof-of-work-based blockchains that exploits the reward mechanism to discourage the miner participation. This BDoS could theoretically be able to grind the (Bitcoin's) blockchain to a halt with significantly fewer resources (21\% of the network's mining power). This attack raises a concern about the liveness of the Proof-of-work-based cryptocurrencies.
This big concern and recent ongoing DoS attack disclosures compel researchers to find new ways to construct efficient DoS mitigation techniques.

%

\subsection{Denial of Service Attack} A denial of service (DoS) attack targets to disrupt the availability of the network, server or application, and prevents legitimate requests from taking place. For a DoS attack to be successful, the attacker has to send more requests than the victim server can handle. These requests can be legitimate or bogus. The DoS attack depletes the server's resources such as CPU, memory, or network. 

\begin{definition}(DoS): Let a server $\mathcal{S}$ be given, with the available resources $R_1, R_2, \\ \ldots, R_n$ ($R_i$ can be bandwidth, memory, CPU etc.). Let $\mathcal{A}$ or a set of $\{\mathcal{A}_j\}$ are an attacker or a set of attackers and let the legitimate users are represented by the set $\{\mathcal{U}_k\}$. A DoS attack on server $\mathcal{S}$ is expressed by a set of probabilities for successful resource-depletion  $\{P_{R_1}, P_{R_2}, \ldots, P_{R_n}  \}$. The total probability for a success of a DoS attack is then a probability the server $\mathcal{S}$ to refuse legitimate transactions from a user $u$, where $u \in \{\mathcal{U}_k\}$ and is modeled as the probability of blocking the legitimate traffic in at least one of the resources:
    \begin{equation} \label{Eq:Probability-DoS}
        P_{DoS} = 1 - (1-P_{R_1})(1-P_{R_2}).\ldots.(1-P_{R_n})
    \end{equation}
\end{definition}
 Note that the situation when attacker(s) exhausts at least one resource $R_i$ implies the attack probability is $P_{R_i} = 1$, which from equation (\ref{Eq:Probability-DoS}) further leads to $P_{DoS} = 1$.

DoS attacks can be categorized into several categories based on network and application layers or volume and protocol attacks. Network-level DoS attacks aim to overload the server's bandwidth or cause CPU usage issues. However, application-level DoS attacks
focus on applications, websites, or online services.

\subsection{Our Contribution}
The contributions of our work are as follows:
\begin{enumerate}
    \item We thoroughly investigate the DoS attacks in the blockchain ecosystem.
    \item We present different mitigation techniques of DoS attacks in the blockchain ecosystem.
    \item We propose a VDF-based DoS resistant protocol by using the functionality of VDF.
\end{enumerate}

The rest of the paper is as follows:  Section~\ref{DoS-Blockchain} shows a detailed analysis of DoS attacks in the blockchain ecosystem.
Further, Section~\ref{Protocol} presents DoS mitigation techniques, including our VDF-based proposal. Finally, in Section~\ref{Conclusion}, we conclude the paper and discuss the possible future directions.

\section{DoS Attacks on Blockchain Ecosystem} \label{DoS-Blockchain}

The blockchain ecosystem has suffered from many DoS attacks in the past, and that situation is a continuing trend. The DoS attack can be launched against a specific entity or a network in the blockchain. We present a nonexclusive list of ten entities in the Blockchain ecosystem with their corresponding DoS attacks. 

\begin{enumerate}[leftmargin=*]

    \item \textit{On cryptocurrency wallets}
    A crypto wallet is a software program in which a user stores cryptocurrency. The wallet contains a set of signing keys for the user to sign new transactions. Wallets are also integrated with decentralized applications (DApps) to hold and manage users' signing keys and transactions securely. In a wallet service, a user is the sole owner of his account keys. However, if someone steals the signing keys, then the cryptocurrency held in that account can be spent. Therefore, hardware wallets (e.g., TREZOR) are ways to store cryptocurrency and the signing key in an offline manner. Nevertheless, online wallets are still a preferable choice for blockchain users. These online crypto wallets also suffer from DoS attacks~\cite{praitheeshan2020security} due to inconsistency in its smart contracts that further hinders the services of integrated DApps. Recently, a DDoS attack was mounted on the Wasabi bitcoin wallet~\cite{Wasabi-DoS}. 
    \item \textit{On cryptocurrency exchange services} A cryptocurrency exchange allows clients to buy, sell and store crypto-currencies at some exchange rate and leverages the clients to trade their currencies and earn some profit due to the fluctuations in the price of currencies. Besides, the exchange charges some fee for every trade made by its client and also converts the cryptocurrency into fiat currencies. Many exchange services also provide a wallet, but the wallet signing keys are controlled by the exchange service apart from the wallet user. Furthermore, these exchange services are online platforms, hence susceptible to DoS attacks that can cause the temporary unavailability of the platform. In the past, many of the crypto-currency exchange services were jeopardized by the DoS attacks (especially DDoS). One such example is the Bitcoin exchange platform, \textit{Bitfinex} which has been a victim of DDoS attacks several times~\cite{Abhishta2019}. Another well-known bitcoin exchange service, Mt. Gox, was completely disrupted by DDoS attacks over time~\cite{feder2017}. Over the years, many cryptocurrency exchange platforms have suffered DoS attacks. Recently, a UK-based exchange \textit{EXMO} was hit by DDoS attack~\cite{EXMO21}.
    

     \item \textit{On memory (transaction) pools} A memory pool (mempool) is a repository of unconfirmed transactions in a cryptocurrency blockchain, e.g., Bitcoin. Once a user creates a new transaction, it is broadcast to the network and stored in the mempool. In the mempool, the transaction waits to be picked by a miner to be added in a block and subsequently to the blockchain, therefore acquiring the transaction's confirmation. If a transaction remains unconfirmed for a long time in the mempool, it gets rejected eventually. As the transactions with high fees are most likely to be selected by a miner, it poses a threat to flood the mempool by small fee transactions, consequently affecting the mempool size. In that direction, it creates uncertainty among the users for their transactions and leads them to pay higher mining fees to prevent the rejection of their transactions. The work~\cite{Saad2018} studies such an attack on Bitcoin mempool and proposes a few countermeasures. However, the proposed solutions have limitations regarding the minimum payable fee and rejection of fast transactions. A follow-up work~\cite{Saad2019} provides similar prevention measures for Proof-of-work-based blockchain but suffers from the similar problems.  
    

    \item \textit{On mining pools} Mining pools are the major players in Proof-of-work-based cryptocurrencies, e.g., Bitcoin. The mining pool's goal is to accumulate miners' power and solve the Proof-of-work puzzles. As the difficulty of Proof-of-work puzzles gives a very low probability of solving the puzzle to a single miner, the miner usually prefers to join a mining pool where the miner gets a fair share of the reward proportional to his/her effort, if the mining pool finds the solution. 
    Two kinds of entities can mount a DDoS attack on a mining pool: 1) A hacker whose aim is to make money by asking the ransom from the attacked mining pool with the promise of stopping the DDoS attack~\cite{MiningPool-DDoS}, 2) A competing mining pool whose goal is to increase his winning probability by undermining the power of competing mining pools. Few game-theoretic studies~\cite{Zheng2019,Wu2020} are also conducted to analyze DoS attacks in mining pools.
    

    

    \item \textit{On layer-two blockchain protocols} Layer-two blockchain protocols are built on the top of the main blockchain that moves a sufficient amount of transaction load from the main blockchain to the off-chain in sub-seconds (instead of minutes or hours) with a reduced fee and similar security. Hence, these protocols are referred to as an orthogonal solution for the scalability problem in the blockchain. In recent years, there has been tremendous growth in constructing new layer-two protocols~\cite{Gudgeon2020} for blockchain scalability such as channel networks. In a channel network, channels are established between the parties of the network and governed by the smart contracts of the main chain. It provides a fast and scalable approach for off-chain interactions. These protocols also suffered from DoS attacks in the past~\cite{tochner2019hijacking,weintraub2020exploiting}.

    \item \textit{On sharding protocols} Similar to layer-2 blockchain solutions, sharding protocols~\cite{wang2019} also tackle the scalability issue of blockchain. The idea of sharding is to partition the blockchain state into multiple shards. Each shard processes a set of transactions; therefore, all shards can process the transactions parallelly and hence increases the blockchain throughput. The majority of the sharding protocols are built on the top of the Bitcoin blockchain, and some are built for the Ethereum blockchain. A sharding protocol deals with challenges involving the shard assignment to validators, transaction assignment to shard, and intra-shard consensus. A DoS attack can be mounted on sharding protocol by flooding a single shard which becomes the bottleneck for the whole system. A recent work~\cite{nguyen2020} studies the DoS-attack on sharding protocols and proposes a Trusted Execution Environment (TEE) based countermeasure.

    \item \textit{On commit-chain operator} A commit-chain~\cite{khalil2018commit} is an off-chain scaling solution where the transactions are performed off-chain by a non-custodial and untrusted operator. The operator commits the balances of users periodically to the blockchain by computing a checkpoint and feeding it to an on-chain smart contract. The scheme involves users publishing challenges to the smart contract in case of a dispute with the operator, which imposes a drawback where a malicious user can flood the smart contract with unwarranted challenges. Another significant issue is the operator being a central entity can become a victim of a DoS attack, resulting in collapsing the whole system. 
    

    \item \textit{On smart contract} A smart contract is a transaction protocol in blockchain that takes actions according to the terms of the contract. In the Ethereum blockchain, each block has a maximum gas limit that is spent by executing a smart contract, and exceeding the gas limit causes a DoS attack. An attacker can mount a DoS attack on smart contract~\cite{Atzei2017} in several possible ways such as: 1) By sending a computationally intensive transaction to a contract thus preventing other transactions from being included in a block; 2) By adding a couple of refund addresses at once that can end up smart contract exceeding the gas limit while refunding to those addresses; 3) By unexpected revert of refund to a legitimate user by using fallback function. A recent work~\cite{samreen2021smartscan} shows a method to detect DoS attacks caused due to unexpected revert in Ethereum smart contract. An example of a DoS attack on a smart contract is an auction contract where an attacker can constantly call the bidding function (e.g., \textit{bid()}), preventing other legitimate users from making their bids. In the NEO blockchain, a vulnerability allowed attackers to invoke a malicious contract that created a DoS attack by crashing each node that tried to execute the contract~\cite{NEO-DoS}. Moreover, a DoS attack on a smart contract triggers stopping a node from executing the functions for all the DApps it hosts.

    \item \textit{On mixing services}
    A mixing service is a protocol that allows a cryptocurrency user to utilize its currency anonymously. It provides unlinkability of the user's input to its output and prevents the user's identification from being revealed. There are centralized~\cite{bao2019lockmix} and decentralized~\cite{ruffing2014} mixing services. Centralized mixing services being a single-point-of-failure are more vulnerable to DoS attacks (e.g. by competing services). However, both types of mixing services suffer from DoS due to different actions of its users, such as 1) By providing inconsistent input for the shuffle, leading the whole verification step of shuffle to fail; 2) By denying to perform some required task e.g., to sign a group transaction; 3) By several participation requests in the mixing transaction pool leading to the depletion of a precomputed pool by participants~\cite{ziegeldorf2018secure}.

    \item \textit{On consensus participants} In the blockchain, consensus participants are the major players who decide on the blockchain's new block. Therefore, consensus participants are the usual DoS target for an attacker. In deterministic leader election protocols of consensus, the leader of the consensus round can be a primary target for DoS attacks which can make the whole system halt if the leader suffers a DoS attack. Other main targets can be stakeholders in Proof of Stake consensus mechanisms that hold some stake in the system, therefore attracting an attacker to mount DoS.
    A DoS attack can be mounted on PBFT-based permissioned blockchains and its participants, where a DDoS attack can be launched if an adversary controls over 33 \% of the replicas. As in the BFT-based blockchains, network size is known to the participants, an attacker creates the required number of Sybil replicas needed for a DoS attack. Hence, for each transaction sent by the primary, the Sybil replicas will not reply to their approvals, leading the whole system to halt. 
    
    
\end{enumerate}

\section{DoS Mitigation Techniques for Blockchain Systems} \label{Protocol}
In most of the DoS events, an attacker floods the network by creating multiple transactions in a short time period, hence maximizing his throughput. This kind of situation arises when the cost of creating a transaction is low. In most settings, these transactions are monetary payment transactions of a tiny value, but for some cases, these can be data transactions (e.g., IoT blockchain transactions). To mitigate the DoS attack, some cost should be imposed on the attacker to slow down or stop unnecessary requests in the blockchain system. Hence, following, we present the DoS mitigation techniques in the blockchain ecosystem.
 \begin{table}[ht]
            \centering
            \begin{tabular}{|l|l|}
            \hline
            \textbf{Blockchain Ecosystem} & \textbf{Applicable Solutions} \\
            \hline
            Cryptocurrency Wallets  &  Client Puzzle (Inside Smart Contract) \\
            \hline
            Cryptocurrency Exchange Services  &  Client Puzzle (On Exchange Clients) \\
            \hline
            Memory Pools    &  Fee-based Approach/NI-Client Puzzle\\
            \hline
            Mining Pools   &  Fee-based Approach/NI-Client Puzzle\\
            \hline
            Layer-2 Blockchain Protocols  &  Fee-based Approach\\
            \hline
            Sharding Protocols   &  Fee-based Approach\\
            \hline
            Commit-chain Operator   &  Client Puzzle (On Commit-chain Users)\\
            \hline
            Smart Contract   &  Client Puzzle (Inside Smart Contract)\\
            \hline
            Mixing Services   &  Fee-based Approach/NI-Client Puzzle\\
            \hline
            Consensus Participants   &  Client Puzzle (On Participant Registration)\\
            \hline
           
            \end{tabular}
            \caption{DoS Mitigation Techniques in Blockchain Ecosystem}
            \label{tab:DoS-mitigation}
\end{table}
 \vspace{-9mm}
\begin{itemize}
    \item \textit{Client Puzzles} Client puzzles are one of the most effective prominent techniques to defend against DoS attacks. In a client puzzle, a client has to solve a puzzle before being granted access to a service or a resource by a server. The initial introduction of the client puzzle was given by Dwork and Naor~\cite{dwork1992} to combat the spam attacks. Client puzzles can be categorized into different types based on the resource used by the client for solving the puzzle such as number of CPU cycles or a number of memory access, quantifying CPU-bound puzzles~\cite{back2002} and memory-bound puzzles~\cite{abadi2005} respectively. Several client puzzles such as Time-lock puzzles~\cite{rivest1996}, Hash-chain~\cite{mahmoody2013} and Equihash~\cite{biryukov2017} are employed in the blockchain ecosystem. A client puzzle scheme can be \textit{Interactive} where server creates the puzzle for the client or \textit{Non-Interactive} (NI) where the client creates a puzzle, solves the puzzle and sends it to the server.
    
    \item \textit{Fee-based Approach} In many events of DoS attack, to disincentivize an attacker an extra or minimum fee can be introduced in the blockchain ecosystem. This fee can be of different types based on the underlying blockchain system. The fee can be a mining fee in mining pools, a mixing fee in mixing services, a transaction fee in transaction pools, a relay fee in a blockchain network, a registration fee for user registration (e.g. a user of a permissioned blockchain), etc. Therefore, with the introduction of a minimum fee, launching a DoS attack becomes costlier for an attacker. However, the fee-based approach adversely affects legitimate users who do not want to pay this minimum amount of fee.  
    
\end{itemize}

Table~\ref{tab:DoS-mitigation} presents the possible DoS mitigation solutions for corresponding blockchain ecosystem. Fee-based approach can be applied in almost every case but will not be favorable for all blockchain users. 
In the table, for layer-2 and sharding protocols, the use of client puzzle will defeat the purpose of scalability due to its time consumption, therefore fee-based approach is a more viable option. For memory pools, mining pools, and mixing services, non-interactive client puzzle schemes can be applied where the miner/user presents a verifiable puzzle and its solution for the inclusion of its new transaction (Rewarding puzzle solution in case of mining pool). Apart from the above described techniques, other mechanisms such as packet filtering techniques or DoS protection services e.g. Incapsula can be used for DoS mitigation in some blockchain contexts.

\subsection{VDF-based DoS-resistant Protocol} 
Most of the existing client puzzles lack public verifiability, non-parallelizability, non-interactivity, and easy verification. Therefore, the initial introduction of VDF~\cite{Boneh2018} as a moderately hard function can be configured as a client puzzle for DoS mitigation achieving all these properties. A VDF can be described as a function $f : \mathcal{X} \rightarrow \mathcal{Y}$ which takes a predefined number of steps $T$ to compute the output $y \in \mathcal{Y}$, given an input  $x \in \mathcal{X}$ and a polynomial number of processors. Furthermore, the verification of the output is exponentially easy. VDF produces a unique output that is efficiently and publicly verifiable. There have been a few constructions of VDF. We employ the Wesolowski VDF scheme~\cite{Wesolowski2019} to construct our client puzzle due to its fast verification and short proof size properties.  

\begin{figure}[th]
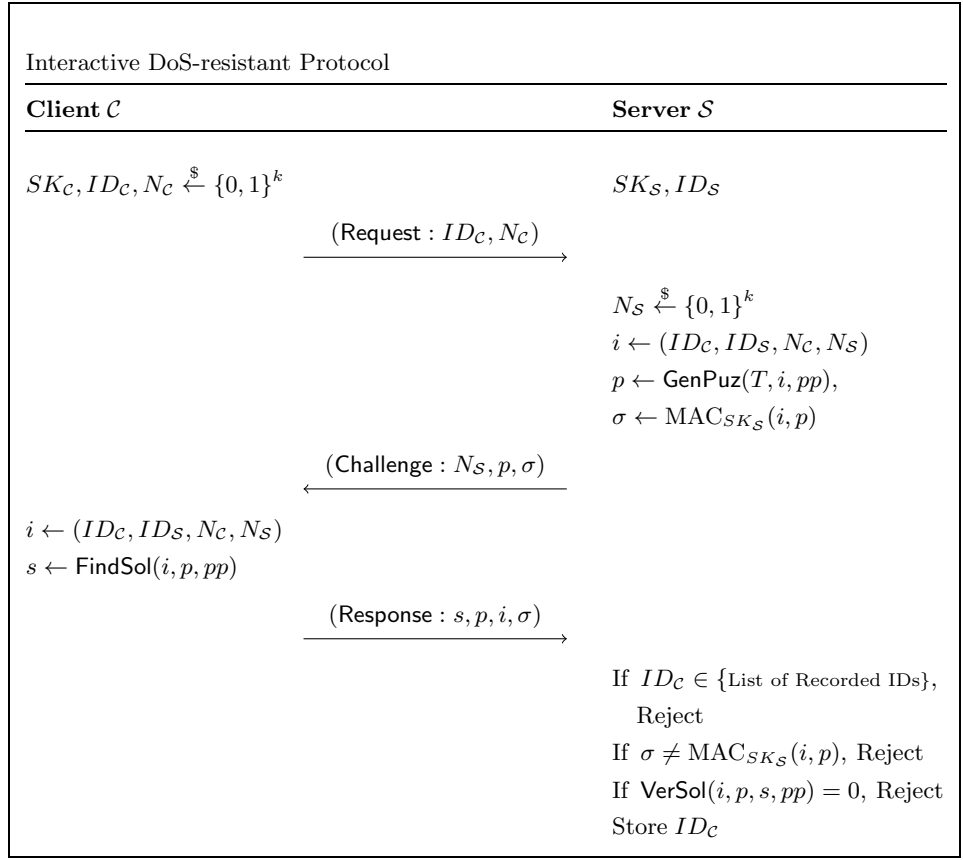

\begin{center}
\fbox{
\parbox{\linewidth}{%
\procedureblock{Interactive DoS-resistant Protocol}{
    \textbf{Client}\; \mathcal{C} \<\< \textbf{Server}\; \mathcal{S} \\[0.7mm][\hline] \\[-1mm]
    SK_\mathcal{C}, ID_\mathcal{C}, N_\mathcal{C} \overset{\$}{\leftarrow} \{0,1\}^k  \<\< SK_\mathcal{S}, ID_\mathcal{S}\\
    \< \sendmessageright*{ (\mathsf{Request}: ID_\mathcal{C}, N_\mathcal{C})} \< \\
    \<\< N_\mathcal{S} \overset{\$}{\leftarrow} \{0,1\}^k \\
    \<\< i \leftarrow (ID_\mathcal{C}, ID_\mathcal{S}, N_\mathcal{C}, N_\mathcal{S}) \\
    \<\< p \leftarrow \mathsf{GenPuz}(T,i,pp), \\
    \<\< \sigma \leftarrow \mathrm{MAC}_{SK_\mathcal{S}}(i,p) \\
    \< \sendmessageleft*{(\mathsf{Challenge}: N_\mathcal{S}, p, \sigma)} \< \\
    i \leftarrow (ID_\mathcal{C}, ID_\mathcal{S}, N_\mathcal{C}, N_\mathcal{S}) \<\< \\
   s \leftarrow \mathsf{FindSol}(i,p,pp) \<\< \\
   \< \sendmessageright*{(\mathsf{Response}: s, p, i, \sigma)} \< \\
   \<\< \text{If}\;\: ID_\mathcal{C} \in \text{\{\scriptsize List of Recorded IDs\}},\: \\
   \<\< \ \ \ \text{Reject} \\
   \<\<  \text{If}\;\: \sigma \neq \mathrm{MAC}_{SK_\mathcal{S}}(i,p),\: \text{Reject} \\
   \<\<  \text{If}\;\: \mathsf{VerSol}(i,p,s,pp) = 0,\: \text{Reject} \\
    \<\< \text{Store}\; ID_\mathcal{C}
}
}
}
\end{center}
\vspace{-3.5mm}
\caption{Interactive DoS-resistant Protocol}\label{Fig:Interactive-DoS-protocol}
\end{figure}

We define an Interactive VDF client puzzle, where a server $\mathcal{S}$ creates a puzzle $p$ and asks for solution $s$ of the puzzle from the client $\mathcal{C}$ before giving access to its resource. In the following construction, $\mathcal{K}$ is a key space, $\mathcal{P}$ is a puzzle space, $\mathcal{O}$ is a solution space, $\mathcal{D}$ is a puzzle difficulty space, and $\mathcal{I}$ is a puzzle input space.
\begin{itemize}
    \item $\mathsf{Setup}(1^{\lambda}$): Select $\mathcal{K} = \varnothing, \mathcal{D} \subseteq \mathbb{N}, \mathcal{P} \subseteq \{0,1\}^*,  \mathcal{O} \subseteq \{0,1\}^*,  \mathcal{I} \subseteq \{0,1\}^*$. Generate a group $\mathbb{G}$ of unknown order, an RSA modulus $N$, a hash function $H:\{0,1\}^* \rightarrow \mathbb{G}$ and $\mathcal{D} \leftarrow T$. Set $param \leftarrow (\mathcal{P, O, D, I})$ and $pp \leftarrow (\mathbb{G}, N, H, T)$, return $(param,pp)$.
    \item $\mathsf{GenPuz}(T,i,pp$): Server runs this algorithm to create a puzzle for the client. It generates an input $i\in \mathcal{I}$ for VDF-evaluation, samples $l \overset{\$}{\leftarrow} Primes(\lambda)$. Return a puzzle $p = l$ to the client.
    \item $\mathsf{FindSol}(i,p,pp$): Client runs this algorithm to solve the puzzle $p$. Client computes $g = H(i)$, further computes $y\leftarrow g^{\left ( 2^{T} \right )} \mathsf{mod}\:N$.
    It computes $q,r$ such that $2^T = ql+r$ where $0 \leq r < l$, and computes a proof $\pi = g^q$. Send a solution $s = (y,\pi)$ to the server.
    \item $\mathsf{VerSol}(i,p,s,pp$): Server computes $r \leftarrow 2^{T} \mathsf{mod}\: l$ and accepts if $g, y, s \in \mathbb{G}$ and $y = {\pi}^{l}{g}^{r} \mathsf{mod}\:N$.
\end{itemize}
An Interactive VDF-based DoS-resistant protocol can be designed using client puzzle as depicted in Figure~\ref{Fig:Interactive-DoS-protocol}. The protocol construction follows from the Stebila et al.~\cite{Stebila2011}. To define this interactive protocol, we assume server and client have public identities $ID_{\mathcal{S}}$ and $ID_{\mathcal{C}}$. Our VDF-based client puzzle can also be made Non-Interactive where the client constructs a puzzle and its solution. The client and server share a common source of randomness (e.g. random beacon). The client creates publicly verifiable puzzles using randomness. Further, the non-interactive VDF client puzzle can be transformed into a DoS-resistant protocol that can be efficiently applied in the blockchain ecosystem during DoS events.

\vspace{-5.5mm}

Following the implementation study of VDF~\cite{attias2020}, for 128-bit security and the difficulty between $2^{16}$ to $2^{20}$, our DoS-resistant protocol can be efficiently employed for DoS mitigation in the blockchain. With the aforementioned setting, the running time for $\mathsf{FindSol}$, $\mathsf{VerSol}$ algorithms are in order of minutes and order of milliseconds respectively. The verification time on the server side can be further optimized using Dimitrov’s multiexponentiation method~\cite{dimitrov2000}. As a future work, we will put a demonstration of a proof-of-concept and initial experiments with Wesolowski VDF for DoS mitigation. 

\section{Conclusion} \label{Conclusion}
In this work, we offered a thorough study of DoS attacks in the blockchain ecosystem. To the best of our knowledge, this is the first investigation in the context of blockchain. As the frequency and intensity of DoS attacks are increasing rapidly, it raises a concern about efficient detection and mitigation techniques. Therefore, we listed out main mitigation approaches which can be used for DoS mitigation in the blockchain ecosystem. We also identify verifiable delay function as an effective primitive to mitigate DoS attacks. A proper construction of non-interactive VDF puzzle and experimental results will be provided in the continuation of this work. This paper will help academic and industrial researchers to study the possible venues and impact of the DoS attack in the blockchain context and to improve upon the existing solutions.

\bibliographystyle{splncs04}
\bibliography{report.bib}

\end{document}